\documentclass[twocolumn]{aa}
\usepackage{graphicx}
\usepackage{txfonts}
\newcommand{\excs}{\extracolsep{\fill}} 
\begin{document}
%
%\headnote{Research Note}
   \title{ Mid-infrared imaging of brown dwarfs in binary systems
\thanks{Based on observations made with ESO Telescopes at the Paranal Observatories under programme ID's 076.C-0556(A) and 077.C-0438(A). }}
%\subtitle{}

   \author{K. Gei{\ss}ler\inst{1,2} \and
   	   G. Chauvin\inst{1,3} \and
	   M. F. Sterzik\inst{1}
          }

   \institute{$^{1}$ European Southern Observatory, Alonso de Cordova 3107, Vitacura, Santiago, Chile\\
              \email{kgeissle@eso.org, gchauvin@eso.org, msterzik@eso.org}\\
  $^{2}$ Max-Planck-Institut f\"ur Astronomie, K\"onigstuhl 17, D-69117 Heidelberg, Germany\\
 $^{3}$  Laboratoire d'Astrophysique, Observatoire de Grenoble, UJF, CNRS; BP 53,
  F-38041 GRENOBLE Cdex 9 (France)\\
             }

   \date{Received <date>; accepted <date>}

   \abstract
	  {Brown dwarfs exhibit complex atmospheric signatures, and their properties depend sensitively on effective temperature, surface gravity, and  metallicity.  Several physical properties of brown dwarfs in binary systems can be well inferred from the primary, and therefore allow to better constrain their atmospheres.}
	    {We want to constrain atmospheric models of brown dwarfs in binary systems using narrow-band mid-infrared photometry.  }
	    {High spatial resolution and sensitivity is required to resolve the components. Therefore we have obtained deep mid-infrared images of four close binary systems with brown dwarf companions using VISIR at the VLT in three narrow-band filters at 8.6, 10.5 and 11.25\,$\mu$m. }
	    {Three brown dwarfs companions (GJ\,229\,B, HD\,130948\,BC and HR\,7329\,B) were detected at 8.6\,$\mu$m. HD\,130948\,BC was also observed at 10.5\,$\mu$m. We finally place upper flux limits for the other narrow band filters with null detections. }
	    {Our results are in general compatible with previous observations and model expectations for these objects. For HD\,130948\,BC, we conclude photometric variability on a significance level of 2.8\,$\sigma$ based on repeated observations.  The bandpass around 10.5\,$\mu$m appears specifically well suited for variability studies, and we speculate that either inhomogeneities in the atmospheric $NH_3$ distribution, or silicate absorption might cause its time-variability.}

   
   \maketitle
%
%
%%%%%%%%%%%%%%%%%%%%%%%%%%%%%%%%%%%%%%%%%%%%%%%%%%%%%%
%
\section{Introduction}

 Brown dwarfs (BDs) bridge the gap in mass between low-mass stars and giant planets. Hundreds of them have been discovered in the past decade, mainly in wide-field optical (SDSS, $\cite{stoughton02}$) and near-infrared (e.g., 2MASS, DENIS; $\cite{cutri03}$, $\cite{epchtein97}$) surveys. Two main classes of BDs emerged from their optical and infrared spectral properties, the  L-dwarfs and  T-dwarfs. They typically cover effective temperature ranges of $\rm{T}_{eff}\,\approx\,2200\,-\,1400\,K$  and $\rm{T}_{eff}\,\approx\,1400\,-\,700\,K$ respectively. The modelisation of atmospheres cooler than $\rm{T}_{\rm{eff}}\leq2000\,K$ is a challenge, because it must include an appropriate treatment of a plethora of molecular opacity's and dust processes (formation, condensation, size distribution and mixing). The most recent atmosphere models are including additional properties such as age (gravity) and metallicity and seem to reproduce reasonably well the spectral signatures and the infrared colors of L and T dwarfs. Only the L-T transition, occurring around a relative narrow temperature range $\rm{T}_{eff}\,\approx\,1300\,-\,1400\,K$, remains problematic (for a discussion of state-of-the-art models see $\cite{burrows06}$).\newline As an additional complication, the determination of distances, ages and metallicity for field BDs remains difficult and may  be the cause of various uncertainties of their physical parameters. In that sense, BD companions of nearby stars are extremely interesting as the system characteristics are well known thanks to the bright primary star. For these systems the influence of age and metallicity on the spectral signatures of substellar objects can be explored with high accuracy.
\begin{table*}[!tb]
\caption[]{Target Properties.}
\label{target}
\begin{tabular*}{\textwidth}{@{\excs}l r r r r r r r r}
%\begin{tabular}{l r r r r r r r r}
\hline
\noalign{\smallskip}
\hline   
\noalign{\smallskip}
Object  & $\pi$ & Age   & [M/H] & separ.   & P.A.     & Spectral & T$_{eff}$  & log\,g  \\
        & [mas] & [Gyr] &       & [arcsec] & [degree] & type B   & [K]        &         \\
\noalign{\smallskip}
\hline  
\noalign{\smallskip}
GJ\,229     (1) & 173.19 & 0.2 (1)             & -0.1\,-\,-0.5 (4) & 7.761\,$\pm$\,0.007 (3) & 163.5\,$\pm$\,0.1 & T6.5                   &  950\,$\pm$\,80 (4)   & 5\,$\pm$\,0.5 (4)  \\
	        &	 & 0.030 (2)           & -0.5 (2)          & 	                     & 	                 &                        & 1000\,$\pm$\,100 (2)  & $\le$\,3.5 (2)  \\
HD\,130948  (5) &  55.73 & 0.3\,-\,0.8 (6-8)   & 0.0 (10)          & 2.64\,$\pm$\,0.01 (5)   & 104.5\,$\pm$\,0.5 & L4\,$\pm$\,1$^\star$(10)& 1900\,$\pm$\,75 (9)  &          \\
HR\,7329   (11) &  20.98 & 0.012 (12)          &                   & 4.17\,$\pm$\,0.05 (11)  & 166.8\,$\pm$\,0.2 & M7/M8 (11)              & 2600\,$\pm$\,200 (11) &               \\
HR\,7672   (13) &  56.60 & 1.\,-\,3.(13)       & 0.02 (10)         & 0.794\,$\pm$\,0.005 (13)& 157.3\,$\pm$\,0.6 & L4.5\,$\pm$\,1.5 (13)   & 1680\,\,$\pm$\,170 (13)&  \\
\noalign{\smallskip}
\hline
\noalign{\smallskip}
\end{tabular*}
\begin{tabular*}{\textwidth}{@{\excs}l}
$^\star$ binary brown dwarf, see text\\
Reference: 1.) $\cite{nakajima95}$; 2.) $\cite{leggett02}$; 3.) $\cite{golimowski98}$; 4.) $\cite{saumon00}$; 5.) $\cite{potter02}$;\\ 6.) $\cite{gaidos98}$; 7.) $\cite{gaidos00}$; 8.) $\cite{fuhrmann04}$; 9.) $\cite{goto02}$; 10.) $\cite{valenti05}$;\\ 11.) $\cite{lowrance00}$; 12.) $\cite{zuckerman01}$ 13.) $\cite{liu02}$
\end{tabular*}
\end{table*}

Initially, most BD studies were conducted in the optical and near-infrared. However, with the recent development of mid-infrared observing techniques and instruments from the ground or from space, this spectral region became essential as numerous molecular lines, including dominant absorption bands of $\rm CH_4, CO, H_2O$ and $\rm NH_3$, are present. The {\it Spitzer Space Telescope} has provided a wealth of high-quality photometry and low- and medium resolution spectroscopy of L and T dwarfs in the 5 -- 20\,$\mu$m regime (see, e.g. $\cite{leggett07}$; $\cite{cushing06}$; $\cite{mainzer07}$). They enable to probe, vertical mixing, clouds and non-equilibrium chemistry around the L -- T boundary in unprecedented detail.

The power of sensitive, ground-based high spatial resolution mid-infrared imaging has recently been demonstrated for the close BD binary companion $\epsilon$\,Indi\,Ba and Bb (see $\cite{sterzik05}$). The {\sl relative photometry} between both components allows to derive effective temperatures,  independently from the determinations in near-infrared. As the distance is well known with the primary, the {\sl absolute photometry} constrains radii and bolometric corrections, in contrast to {\it Spitzer} observations  ($\cite{roellig04}$) that suffered from insufficient angular resolution to resolve both components ({\it Spitzer} diffraction limit at 10\,$\mu$m is $\sim$\,3\arcsec). 

In order to continue our effort to constrain atmospheric and evolutionary models of BDs, we have therefore conceived a mini-survey of close BD companions using the mid-infrared imager VISIR at the VLT. Section 2 reviews the targets properties and describes our observations. Section 3 describes our data reduction and analysis process. Section 4 presents the results of our survey, i.e the photometry and the astrometry of the detected companions and the sensitivity limits obtained. Finally, Section 5 compares our results with the predictions of cool atmosphere models to discuss their applicability and limitations.
%%%%%%%%%%%%%%%%%%%%%%%%%%%%%%%%%%%%%%%%%%%%%%%%%%%%%%
%%%%%%%%%%%%%%%%%%%%%%%%%%%%%%%%%%%%%%%%%%%%%%%%%%%%%%
\section{Target Properties and Observation}
For our target selection, we have only considered the confirmed members of binary (or multiple) systems with known distances, as their primaries are well characterized in terms of metallicity and age. Then, only BD companions with expected mid-IR fluxes \textbf{stronger than 1\,mJy} and separations larger then 0.5\arcsec were selected in order to fully adapt and exploit the sensitivity and spatial resolution of the mid-infrared imager VISIR at the VLT. We finally ended-up with a short list of four systems: GJ\,229, HD\,130948, HR\,7329 and HR\,7672. Their main characteristics are summarized in Table\,$\ref{target}$. \\ 
\begin{table*}[tb]
\caption[]{Observing Log.}
\label{log}
\begin{tabular*}{\textwidth}{@{\excs}l r r r r r r r r r l}
\hline
\noalign{\smallskip}
\hline   
\noalign{\smallskip}
Object &  Filter & UT date & Airmass & DIMM$^{a}$       & humidity  & DIT & NDIT & \# of       & Int. Time$^{d}$ & Calibrator \\
       &	 & dd/mm/yr&         & seeing [\arcsec] & [$\%$]    & [s] &      & nods$^{c}$  &  [s]            & \\
\hline  
\noalign{\smallskip}
GJ\,229 & PAH1  & 08/01/06$^{b}$ & 1.109 & 0.91 & 17\,-\,50 & 0.016 & 123 & 12 & 2172.7 & HD\,26967, HD\,75691\\
	&	& 10/02/06$^{b}$ & 1.195 & 0.79 & 23\,-\,55 & 0.016 & 123 & 18 & 3259.0 & HD\,41047, HD\,75691\\
	& PAH2 	& 03/02/06$^{b}$ & 1.044 & 0.83 & 20\,-\,40 & 0.008 & 246 &  6 & 1086.3 & HD\,41047, HD\,26967\\
	&	& 10/02/06       & 1.110 & 0.84 & 23\,-\,55 & 0.008 & 246 &  6 & 1086.3 & HD\,26967, HD\,75691\\
	& SIV 	& 10/02/06       & 1.022 & 0.95 & 23\,-\,55 & 0.04  &  48 & 11 & 1943.0 & HD\,26967, HD\,75691\\
\noalign{\smallskip}
\hline
\noalign{\smallskip}
HD\,130948 & PAH1 & 09-10/07/06       & 1.524 & 0.67 & 6\,-\,8  & 0.020 & 98 & 22 & 3967.0 & HD\,133774, HD\,99167\\
	   & PAH2 & 11/07/06$^{b}$    & 1.541 & 0.91 & 7\,-\,10 & 0.010 & 197& 22 & 3987.3 & HD\,149009\\
	   &	  & 04-05/08/06$^{b}$ & 1.734 & 0.71 & 7        & 0.010 & 197& 11 & 1993.6 & HD\,149009, HD\,145897\\
	   & SIV  & 03/08/06          & 1.627 & 0.86 & 5\,-\,8  & 0.040 & 48 & 11 & 1943.0 & HD\,145897\\
	   &	  & 05/08/06          & 1.606 & 0.62 & 6\,-\,11 & 0.040 & 48 & 11 & 1943.0 & HD\,145897\\
\noalign{\smallskip}
\hline
\noalign{\smallskip}
HR\,7329 & PAH1	& 07/06/06 & 1.182 & 1.05 & 4\,-\,12 & 0.020 & 98 & 22 & 3967.0 & HD\,178345\\
	 & PAH2	& 21/05/06 & 1.153 & 1.36 & 7\,-\,19 & 0.008 & 246& 11 & 1991.6 & HD\,178345\\
	 & SIV	& 07/06/06 & 1.175 & 0.88 & 4\,-\,12 & 0.040 & 48 & 22 & 3886.1 & HD\,178345\\
\noalign{\smallskip}
\hline
\noalign{\smallskip}
HR\,7672 & PAH1	& 10/07/06$^{b}$ & 1.341 & 0.76 & 6\,-\,8  & 0.020 & 98 & 11 & 1983.5 & HD\,189695, HD\,149009\\
	 &	& 12/07/06$^{b}$ & 1.405 & 0.88 & 6\,-\,11 & 0.016 & 123& 11 & 1991.6 & HD\,189695, HD\,178345\\
	 & PAH2	& 10/07/06$^{b}$ & 1.385 & 0.81 & 6\,-\,8  & 0.010 & 197& 11 & 1993.6 & HD\,189695, HD\,220954\\
	 & 	& 13/07/06$^{b}$ & 1.693 & 0.78 & 5\,-\,8  & 0.010 & 197& 11 & 1993.6 & HD\,198048, HD\,217902\\
	 & SIV 	& 13/07/06$^{b}$ & 1.356 & 0.82 & 5\,-\,8  & 0.040 & 48 & 15 & 2649.6 & HD\,178345, HD\,198048\\
\noalign{\smallskip}
\hline
\end{tabular*}
\begin{tabular*}{\textwidth}{@{\excs}l }
a.) in V band at 550\,nm, b.) data showed stripes, c.) 23 chops per nod, d.) t\,=\,DIT\,$\times$\,NDIT\,$\times$\,\# of chops\,$\times$\,\# of nods\,$\times$\,4
\end{tabular*}
\end{table*}
\begin{itemize}
\item GJ\,229\,B is the first unambiguous discovered BD ($\cite{nakajima95}$). Later on, orbital motion was detected by $\cite{golimowski98}$, who observed GJ\,229\,B at three epochs spread over approximately one year using HST's Wide Field Planetary Camera 2 (WFPC2). $\cite{matthews96}$ derived an effective temperature of $\sim$\,900K from the measured broadband spectrum of GJ\,229\,B, assuming a radius equal to that of Jupiter. The same effective temperature was obtained by $\cite{leggett99}$ by comparing colours and luminosity to evolutionary models by $\cite{burrows97}$. In general model spectra for GJ\,229\,B ($\cite{marley96}$; $\cite{allard96}$) reproduce the overall energy distribution fairly well and agree with $\rm T_{eff}$\,=\,950\,K.\\
\item HD\,130948\,BC is a binary brown dwarf companion detected by $\cite{potter02}$. The separation between the two companions is (0.134\,$\pm$\,0.005)\arcsec at PA\,=\,(317\,$\pm$\,1)\degr. Both companions have the same spectral type (L4\,$\pm$\,1,) and effective temperatures ($\rm T_{eff}$\,=\,1900\,$\pm$\,75\,K) ($\cite{goto02}$).\\ 
\item HR\,7329\,B is a BD companion detected by $\cite{lowrance00}$ at a separation of 4\arcsec\, from the early-type star HR\,7329\,A, a member of the  $\beta$ Pictoris moving group ($\cite{zuckerman01}$). Their optical spectrum points toward a spectral type M7/M8 and an effective temperature of 2405 to 2770\,K for this young substellar companion. $\cite{guenther01}$ show evidence that the source is a co-moving companion.\\
\item HR\,7672\,B, a common proper motion companion to the variable star HR\,7672\,A, was reported by $\cite{liu02}$. They inferred an effective temperature of T$_{eff}$\,=\,1510\,-\,1850\,K for HR\,7672\,B and estimated an age of 1\,-\,3\,Gyr for the system.\\
\end{itemize}
All targets have been observed using  VISIR ($\cite{lagage04}$) mounted at the UT3 (Melipal) with the filters PAH1 ($\lambda_{cen}$\,=\,8.59\,$\mu$m, $\Delta\,\lambda$\,=\,0.42\,$\mu$m), PAH2 ($\lambda_{cen}$\,=\,11.25\,$\mu$m, $\Delta\,\lambda$\,=\,0.59\,$\mu$m) and SIV ($\lambda_{cen}$\,=\,10.49\,$\mu$m, $\Delta\,\lambda$\,=\,0.16\,$\mu$m). A nominal pixelscale of 0.075\arcsec\, was used during all observations and standard chopping and nodding techniques, with a chop-throw amplitude of 6\arcsec\, and 8\arcsec\, for GJ\,229, respectively, and a chopping frequency of 0.25 Hz were employed. The nodding direction was chosen parallel to the chopping direction and, consequently, with an equal nodding to chopping amplitude.  \newline To assure that the primary as well as the companion are within the FoV during chopping and nodding the system was aligned horizontally on the detector. At the same time, this simplifies the reduction process, since the shift and add of the frames can be done using the brighter primary star. A summary of the observing log is given in Table\,$\ref{log}$., including the mean airmass during the observing run  and the total integration time.
%%%%%%%%%%%%%%%%%%%%%%%%%%%%%%%%%%%%%%%%%%%%%%%%%%%%%%%
%%%%%%%%%%%%%%%%%%%%%%%%%%%%%%%%%%%%%%%%%%%%%%%%%%%%%%%
\section{Data Analysis}
\subsection{Data Reduction}
The reduction was done using self-written IDL routines for bad-pixel replacement as well as for the shift and add of the frames. Bad-pixel were replaced by the mean value of the surrounding pixels within a box of 9\,$\times$\,9 pixel, before subtracting (A-B) nodding positions. In the following the relative shifts between the frames of one data-set were calculated via cross-correlation of the bright primary, before the frames were averaged. Since the VISIR detector was affected by randomly triggered stripes during part of the observations, a destriping technique developed by $\cite{pantin07}$ was applied to the final, co-added images. \\
%%%%%%%%%%%%%%%%%%%%%
% FIGURE 1
%%%%%%%%%%%%%%%%%%%%%
\begin{figure}[tb!]
%\begin{center}
\resizebox{\hsize}{!}{\includegraphics[angle=90]{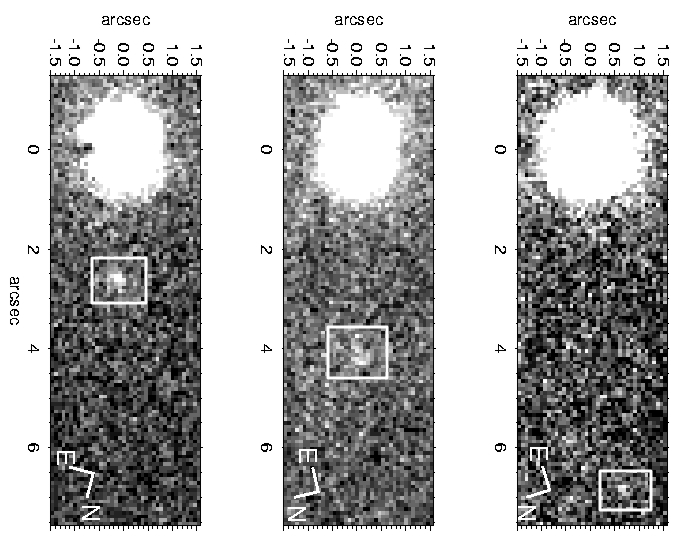}}
%\end{center}
\caption{VISIR detection images of GJ\,299B at $\sim$\,6.8\arcsec\, (upper), HR\,7329\, at $\sim$\,4.3\arcsec\, (middle) and HD\,130948\,BC at $\sim$\,2.5\arcsec\, (bottom) at 8.6\,$\mu$m. To all images a $\sigma$ filter with a box width of 5 pixel has been applied. Furthermore the N-E orientation of the data is over plotted in the lower right corner of the image.}
\label{pic_pah1}
\end{figure}
%%%%%%%%%%%%%%
\subsection{Aperture Photometry}
%%%
Standard aperture photometry was used to determine the relative photometry of all detected BDs. Using IDL ATV routines, curve-of-growth method was applied to the brown dwarf companions to obtain the apertures where the signal-to-noise ratio is maximised. In the following, those apertures were used for the primary as well as for the standard stars. Thus the count-rate to flux conversion factor was determined and relative photometry obtained. The variation of the count-rate to flux conversion factors with aperture radius was screened for at least 3 consecutive aperture radii between 4 and 7 pixels (corresponding to radii of 0.3\arcsec\, to 0.525\arcsec). At 10\,$\mu$m the VISIR diffraction limit is $\sim$\,0.3\arcsec, the chosen aperture radii are of the order or twice the diffraction limit. To calibrate the flux values different standard stars,\protect\footnote{Taken from the list of $\cite{cohen99}$} observed before and after the targets, were used. The error bar estimation of the flux calibration is derived from the flux variations of the source measured in different apertures and from different standard stars. In case two independent measurements were taken an average of the measured flux is quoted in Table\,$\ref{phot}$.

As already mentioned a destriping technique was applied to the final images to clean it from random stripes and therewith improve the image quality. To estimate the impact of the destriping on the photometry of the BD's, we performed the aperture photometry before and after the destriping process. In those cases in which the source is not located close to a stripe, no influence is notable. Whereas in the cases where the BD is close-by a stripe, a decrease in the measured count-rate, and consequently in flux, of the BD in the destriped images is perceivable. Nevertheless this effect is expected, since the stripes in the 'raw' images fall within the aperture radii and lead to an overestimation of the count-rate and therewith of the flux.
%%%%%%%%%%%%%%%%%%%%%%%%
\subsection{Detection limits}
%%%%%
To estimate the detection limits as a function of angular separation two approaches were explored. The standard deviation of the intensities within a 1 pixel wide annulus at a given radius was determined as well as the standard deviation within a box of 5\,$\times$\,5 pixels along a random radial direction. From the obtained noise estimate the contrast with respect to the peak intensity of the primary was calculated (see, e.g., Figure\,$\ref{det_siv}$). The detection limits delivered by both methods are in good agreement. Additionally, to further test the derived detection limits, artificial companions, with fluxes varying between 2 and 10\,mJy, were placed within the data at separation between 1 and 5\arcsec. The limiting fluxes of the re-detectable artificial companions match the previously derived detection limits. Up to a separation of $\sim$\,1.5\,\arcsec\, the detection limit is dominated by the photon noise of the central star and at larger separation the background noise from the atmosphere and the instrument limits our detections.
%%%%%%%%%%%%%%%%%%%%%
%%%%%%%%%%%%%%%%%%%%%
% FIGURE 2
%%%%%%%%%%%%%%%%%%%%%
\begin{figure*}[!tb]
\begin{center}
\resizebox{\hsize}{!}{\includegraphics[angle=90]{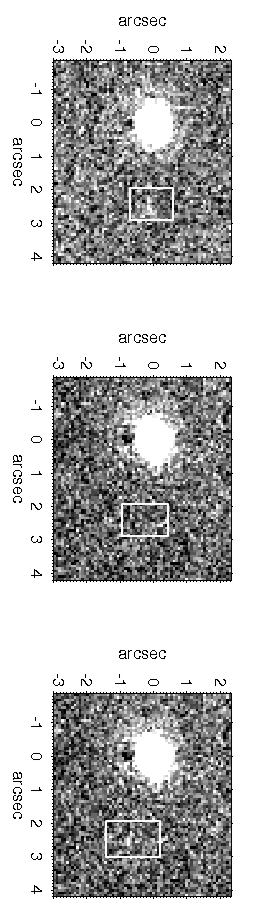}}
\end{center}
\caption{VISIR images of HD\,130948 in the SIV filter taken on the 5th of August (left) and on the 3rd of August (middle). The binary brown dwarf companion was detected in the data from the 5th of August and is marked by a box in the leftmost image. The flux was measured to be 5.7\,$\pm$\,0.4\,mJy. In the data set from the 3rd of August the companion was not detected. Its approximate location is at the same position as in the left image and also marked by a box. The right image shows the data from the 3rd of August, in which an artificial companion of 4\,mJy had been placed. The artificial companion is located somewhat below the expected position of the real companion.}
\label{pic_siv}
\end{figure*}
%%%%%%%%%%%%%%%%%%%
%%%%%%%%%%%%%%%%%%%
\section{Results}
Three of the four brown dwarfs were detected in PAH1, namely GJ\,229\,B, HR\,7329\,B and HD\,130948\,BC (see Figure\,$\ref{pic_pah1}$); while only HD\,130948\,BC could be detected in SIV. Note that HD\,130948\,BC, a binary brown dwarf, was not resolved in our observations. In none of the filters HR\,7672\,B could be detected. While the resolution of VISIR is sufficient to separate the brown dwarf and the primary (($\sim$\,0.79)\arcsec\, $\cite{liu02}$; assuming orbital motion negligible), the data quality in PAH1 and SIV is low. The PSF of the primary is elongated, affecting the area in which the brown dwarf is expected, and thus adding noise.

\begin{figure}[!tb]
\begin{center}
\resizebox{\hsize}{!}{\includegraphics[angle=90]{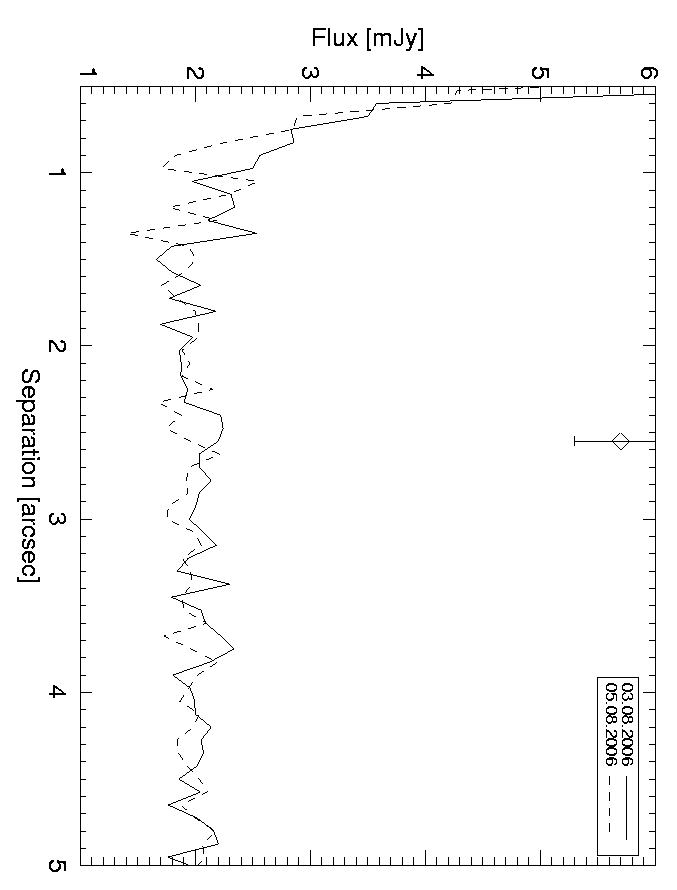}}
\end{center}
\caption{Comparison of the limiting background obtained from the two data sets of HD\,130948 taken in the SIV filter. The over-plotted point corresponds to the detection on the 5th of August, with a flux of 5.7\,mJy.}
\label{det_siv}
\end{figure}

In Table\,$\ref{sep}$. the measured separations and position angles of the detected brown dwarfs are given. To obtain the separation as well as the position angle of the brown dwarfs relative to their primaries, the pixelscale and N-orientation provided in the image header were used. $\cite{golimowski98}$, used the HST's Wide Field Planetary Camera 2 (WFPC2) to observe GJ\,229\,B at three epochs, which were spread over approximately one year. Orbital motion of GJ\,229\,B was detected and a relative change of separation of (0.088\,$\pm$\,0.010)\arcsec\, per year was measured. In the last 10 years, from November 1996 to February 2006, the separation between GJ\,229\,A and B changed by (0.894\,$\pm$\,0.05)\arcsec, resulting in a average change of separation of (0.097\,$\pm$\,0.005)\arcsec\, per year. For HD\,130948 only a minor change of separation is observable. From February 2001 to July 2006 the separation between HD\,130948\,A and BC declined by (0.09\,$\pm$\,0.05)\arcsec. In case of HR\,7329 no orbital motion was observable.
%%%%%%%%%%%%%%%
%%%%%%
\begin{table}[htb!]
\caption{Separations and position angles of the detected brown dwarfs.}
\label{sep}
\begin{tabular*}{\columnwidth}{@{\excs}l r r r}
\hline
\noalign{\smallskip}
\hline   
\noalign{\smallskip}
Object & UT date &  sep. [arcsec] & P.A. [$\degr$] \\
\noalign{\smallskip}
\hline
\noalign{\smallskip}
GJ\,229    & 10/02/06 & 6.78\,$\pm$\,0.05 & 168.4\,$\pm$\,0.9 \\
\noalign{\smallskip}
HD\,130948 & 09/07/06 & 2.54\,$\pm$\,0.05 & 103.9\,$\pm$\,2.4 \\
\noalign{\smallskip}
HR\,7329   & 07/06/06 & 4.17\,$\pm$\,0.11 & 167.2\,$\pm$\,1.4 \\
\noalign{\smallskip}
\hline
\end{tabular*}
\end{table}
%%%%%%%%
%%%%%%%%%%%%%%%%%%%%%%%%%%%%%%%%%%%%%%%%%%%%%%%%%%%
Finally in Table\,$\ref{phot}$. the obtained fluxes for the primary stars and the brown dwarfs, as well as the upper limits for the non-detections are listed. In case of HD\,130948\,BC the flux measured in the data set from the 5th of August is quoted as well as the upper limit obtained on the 3rd of August. While the observations of HD\,130948\,BC in SIV have been carried out at two different epochs, on the 3rd and the 5th of August, the object was only detectable in the second data set, see Figure\,$\ref{pic_siv}$ with a measured flux of (5.7\,$\pm$\,0.4)\,mJy. The non-detection of HD\,130948\,BC in the data set from the 3rd of August can not be explained by a discrepancy in the sensitivity limits, see Figure\,$\ref{det_siv}$. Both data sets clearly reach the same sensitivity limit. Furthermore, simulations of artificial sources showed, that a companion with a flux of (4\,$\pm$\,0.4)\,mJy (corresponding to a 5\,$\sigma$ confidence level) would have been detected in both data sets. Hence, within $\approx$\,48\,hours HD\,130948\,BC varied by at least (1.7\,$\pm$\,0.6)\,mJy. 
%%%%%%%%%%%%%%%%%%%%%%%%%%%%%%%%%%%%%%%%%%%%%%%%%%%%%
%%%%%%
\begin{table}[tb]
\caption{VISIR Photometry of the primaries and brown dwarfs. In case of a non-detection upper limits are provided. The fluxes are quoted in mJy.  }
\label{phot}
{\small
\begin{tabular*}{\columnwidth}{@{\excs}l r r r}
\hline
\noalign{\smallskip}
\hline   
\noalign{\smallskip}
Object &  PAH1 & SIV & PAH2 \\
\noalign{\smallskip}
\hline
\noalign{\smallskip}
GJ\,229\,A    & 1297. (47) & 923. (33.)         & 793. (26.) \\
GJ\,229\,B    & 3.2  (0.5) & 2.   (0.3)$^{a}$   & 4.$^{a}$ (0.9) \\
\noalign{\smallskip}
%HD\,130948\,A & 861. (5.) & 580. (38.)         & 478. (6.) \\
HD\,130948\,A  & 861. (5.) & 605. (27.)$^{b}$   & 478. (6.) \\
HD\,130948\,BC & 3.8 (0.4) & 5.7  (0.4)$^{b}$   & 1.8 (0.2)$^{a}$ \\
HD\,130948\,A  &           & 553. (27.)$^{c}$   & \\
HD\,130948\,BC &	   & 2.   (0.4)$^{a,c}$ & \\
\noalign{\smallskip}
HR\,7329\,A   & 524. (19.) & 404. (3.)          & 386. (24.) \\
HR\,7329\,B   & 3.2  (2.3) & 1.3  (0.2)$^{a}$   & 2.3 (0.2)$^{a}$ \\
\noalign{\smallskip}
HR\,7672\,A   & 880. (36.) & 554. (33.)         & 519. (14.) \\
%HR\,7672\,B  & 5.3$^{a}$  & 6.1$^{a}$          & 3.$^{a}$ \\
\noalign{\smallskip}
\hline
\noalign{\smallskip}
\end{tabular*}
}
\begin{tabular*}{\textwidth}{@{\excs}l }
a.) limiting background (1\,$\sigma$), b.) only 05.08.2006, c.) 03.08.2006
\end{tabular*}
\end{table}
%%%%%%
%%%%%%%%%%%%%%%%%%%%%%%%%%%%%%%%%%%%%%%%%%%%%%%%%%%%%
\section{Discussion}
\subsection{Comparison with models}
As a final step we compare our obtained photometry to the models by $\cite{allard01}$ and $\cite{burrows06}$. Using their theoretical spectra provided online\protect\footnote{http://perso.ens-lyon.fr/france.allard/ and \hspace{2cm} http://zenith.as.arizona.edu/burrows/} absolute model fluxes were calculated by integrating the theoretical spectrum over the VISIR filter bandpasses. The object radii R, which determine the absolute spectral flux calibration, were obtained from evolutionary calculations by $\cite{burrows97}$. In Table\,$\ref{model}$. the calculated model fluxes are listed. Furthermore the age and effective temperature combinations for which the object radius was determined are given.
\begin{table*}[tb]
\caption{Predicted mid-IR fluxes from different theoretical models. Values consistent within 3\,$\sigma$ or with the given upper limit were marked in {\bf boldface}. }
\label{model}
\begin{center}
\begin{tabular*}{\textwidth}{@{\excs}l l r r r r r r r r}
\hline
\noalign{\smallskip}
\hline   
\noalign{\smallskip}
Object  & Reference & T$_{eff}$ & log\,g     & [M/H] & age & R/R$_{\odot}$ & PAH1  &  SIV  & PAH2 \\
	&	    &  [K]      & [cm/s$^2$] &       & Myr &               & [mJy] & [mJy] & [mJy]\\
\noalign{\smallskip}
\hline
\noalign{\smallskip}
GJ\,229\,B & Allard$^a$ &  900 & 5.0 &  0.0 & 200 & 0.122 & {\bf 3.30}      & {\bf 2.97} & {\bf 5.06} \\
	   & Allard$^a$ & 1000 & 5.0 &  0.0 & 200 & 0.122 & {\bf 4.67}      & 4.55       & {\bf 6.68} \\
	   & Allard$^a$ & 1000 & 3.0 &  0.0 & 200 & 0.122 &  4.80           & 6.69       & 7.94 \\
	   & Allard$^a$ & 1000 & 3.0 &  0.0 &  30 & 0.133 &  5.70           & 7.95       & 9.44 \\ 
	   & Burrows$^c$&  900 & 5.0 &  0.0 & 200 & 0.122 & {\bf 3.35}      & {\bf 3.0}  & {\bf 4.25} \\
	   & Burrows$^c$&  900 & 5.0 & -0.5 & 200 & 0.122 & {\bf 3.33}      & {\bf 2.46} & {\bf 3.78} \\
	   & Burrows$^c$& 1000 & 5.0 & -0.5 & 200 & 0.122 & {\bf 4.25}      & 3.44       & {\bf 4.77} \\
	   & Burrows$^c$& 1000 & 5.0 & -0.5 &  30 & 0.133 &  5.04           & 4.09       & {\bf 5.66} \\
	   & Burrows$^c$& 1000 & 4.5 & -0.5 &  30 & 0.133 &  5.14           & 4.99       & {\bf 6.05} \\
           & measured   &      &     &      &     &       & 3.2\,$\pm$\,0.5 & 3.2$^1$    & 6.7$^1$\\
\noalign{\smallskip}
\hline
\noalign{\smallskip}
HD\,130948\,BC & Allard$^b$  & 1900 & 5.0 & 0.0  & 800 & 0.091 & {\bf 1.65}      & {\bf 1.25} & {\bf 1.13} \\
	       & Allard$^b$  & 1900 & 5.0 & 0.0  & 300 & 0.102 & {\bf 2.07}      & 1.57       & 1.42 \\
 	       & Allard$^b$  & 1900 & 3.5 & 0.0  & 300 & 0.102 & {\bf 1.67}      & {\bf 1.21} & {\bf 1.11} \\
	       & Burrows$^d$ & 1900 & 5.0 & 0.0  & 300 & 0.102 & {\bf 1.53}      & {\bf 1.26} & {\bf 1.17} \\
	       & Burrows$^d$ & 1900 & 5.0 & -0.5 & 300 & 0.102 & {\bf 1.49}      & {\bf 1.25} & {\bf 1.16} \\
               & measured$^2$&      &     &      &     &       & 1.9\,$\pm$\,0.3 & 2.9\,$\pm$\,0.3$^3$ & 1.2$^1$\\
               & measured$^2$&      &     &      &     &       &                 & 1.5$^1$              &    \\
\noalign{\smallskip}
\hline
\noalign{\smallskip}
HR\,7329\,B & Allard$^b$  & 2400 & 4.0 & 0.0 &  12 & 0.193 & {\bf 1.06}      & {\bf 0.91} & {\bf 0.79} \\
	    & Allard$^b$  & 2600 & 3.5 & 0.0 &   8 & 0.278 & {\bf 2.80}      & 2.30       & {\bf 2.55} \\
	    & Allard$^b$  & 2600 & 3.5 & 0.0 &  12 & 0.229 & {\bf 1.90}      & {\bf 1.56} & {\bf 1.73} \\
	    & Allard$^b$  & 2600 & 4.0 & 0.0 &  12 & 0.229 & {\bf 1.76}      & {\bf 1.44} & {\bf 1.25} \\
	    & Allard$^b$  & 2800 & 4.0 & 0.0 &  12 & 0.265 & {\bf 2.84}      & 2.20       & {\bf 1.91} \\
            & measured    &      &     &     &     &       & 3.2\,$\pm$\,2.3 & 1.9$^1$    & 2.9$^1$\\
\noalign{\smallskip}
\hline
\noalign{\smallskip}
\end{tabular*}
\end{center}
\begin{tabular*}{\textwidth}{@{\excs}l l}
a.) AMES-cond models from $\cite{allard01}$ & 1.) upper limit, limiting background plus 3$\sigma$\\
b.) AMES-dusty models from $\cite{allard01}$ & 2.) flux of one component assuming both L4 dwarfs contribute equally \\
c.) L \& T cloud-free model from $\cite{burrows06}$ & \hspace{0.35cm} to the measured flux \\
d.) L \& T model with clouds from $\cite{burrows06}$ & 3.) only 05.08.2006\\
\end{tabular*}
\end{table*}
From $\cite{allard01}$ we employed the AMES-cond and AMES-dusty models, representing the two extreme cases, in which either all dust has disappeared from the atmosphere (AMES-cond) or dust settling throughout the atmosphere is negligible (AMES-dusty). Following $\cite{allard01}$ the AMES-dusty models should successfully describe dwarfs with effective temperatures greater than 1800\,K, while the AMES-cond models are better suited to describe the atmospheres of dwarfs with T$_{eff}$\,$\le$\,1300\,K.
%%%%%%%%
%%%%%%%%%%%%%%%%%%%%%%%%%%%%

\begin{figure}[tb]
\begin{center}
\resizebox{\hsize}{!}{\includegraphics[angle=90]{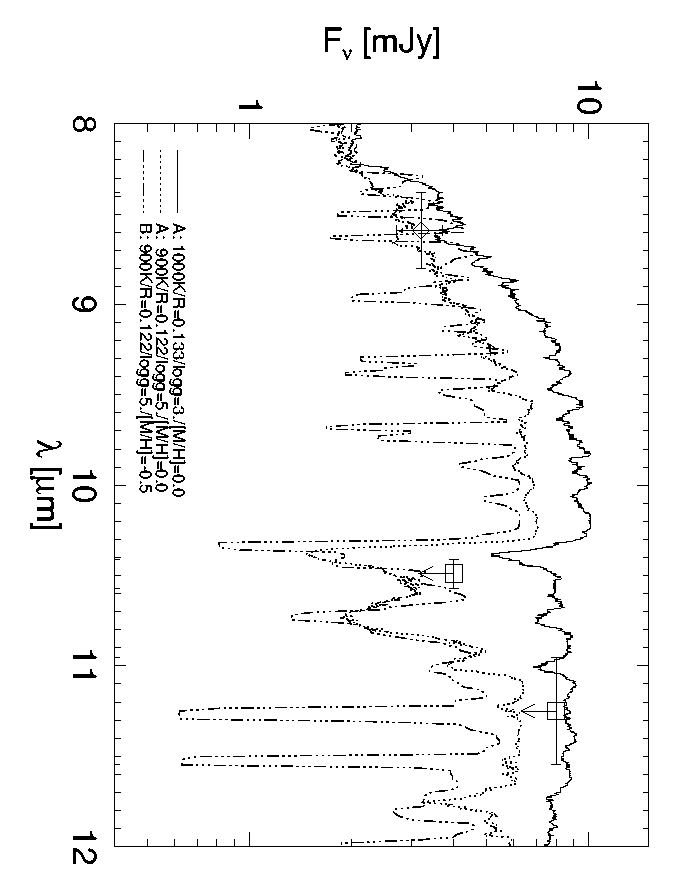}}
\end{center}
\caption{GJ\,229\,B: Comparison of theoretical spectra from $\cite{allard01}$ and $\cite{burrows06}$ with the VISIR photometry.}
\label{spec}
\end{figure}

{\bf GJ\,229\,B:}
$\cite{saumon00}$, have used high-resolution infrared spectra to determine the metallicity, effective temperature and gravity of the T dwarf (see also Table\,$\ref{target}$.). While using an age of 0.2 Gyr they derived an effective temperature of 950\,K\,$\pm$\,80\,K and a gravity of log\,{\it g}\,=\,5\,$\pm$\,0.5. Later on, $\cite{leggett02}$ compared observed low- and high- resolution spectra of GJ\,229\,A and GJ\,229\,B to theoretical spectra (AMES-models). Their best fit yields an $\rm T_{eff}$\,=\,1000\,$\pm$\,100 and a gravity of log\,{\it g}\,$\le$\,3.5 for GJ\,229\,B as well as a metallicity of [M/H]\,$\approx$\,-0.5 and an age of $\sim$\,30\,Myr (range 16\,-\,45\,Myr) for the system. While the metallicity is determined within the spectra fitting procedure, the age is derived by a comparison with evolutionary models and mainly constrained by the observed luminosity and the derived effective temperature of the A component. Using VISIR mid-IR photometry, absolute model predictions of both $\cite{allard01}$ and $\cite{burrows06}$ can then be tested for different combinations of Teff, [M/H] and gravity (when available, e.g. see Figure/,$\ref{spec}$). From Table\,$\ref{model}$, we see that the PAH1, SIV and PAH2 photometry is consistent with model predictions for a $\rm T_{eff}$\,=\,900\,K, log\,{\it g}\,=\,5.0 and [M/H]\,=\,0 companion. At solar metallicity, an effective temperature of $\rm T_{eff}$\,=\,1000\,K can be excluded at more than 2 sigma. At subsolar [M/H]\,=\,-0.5 metallicity, low gravity (log\,{\it g}\,$<$\,4.5) values remain excluded for $\rm T_{eff}$\,=\,1000\,K. Therefore, excluding young ages predictions of $\cite{leggett02}$, the VISIR photometry clearly favors the initial physical parameters proposed by $\cite{saumon00}$ for solar and subsolar metallicities.
%%%%%%%%%%%%%%%%%%%
%%%%%%%%%%%%%%%%%%%%%%%%%%%%

{\bf HD\,130948\,BC:} As already mentioned HD\,130948\,BC is a binary brown dwarf consisting of two L\,4 dwarfs. In order to compare our observations to the model predictions we assumed that both brown dwarfs contribute equally to the measured fluxes,  as simplest assumption. An unequal distribution of the measured flux on one of the two brown dwarfs would only increase the in the following described effect. The suggested effective temperature of $\sim$\,1900\,K places the binary in the regime of the AMES-dusty models. While the predicted PAH1 flux is in good agreement with our measurements, most models fail to reproduce the SIV flux detected on August 5 2006, but would be consistent with the non-detection of August 3 2007.  Using the models provided by $\cite{burrows06}$ we tested the influence of different metallicities ([M/H]=0.0 and [M/H]=-0.5) on the predicted fluxes. The change is marginally, only about 0.05\,mJy.
%%%%%%%%%%%%%%%%%%%
%%%%%%%%%%%%%%%%%%%%%%%%%%%%
\subsection{HD\,130948\,BC: photometric variability} 
While analysing the SIV data obtained for HD\,130948, we found that the binary companion was detectable in only one of the two data sets. A possible explanation of this result is an intrinsic variability around 10.5\,$\mu$m of one or both of the L dwarfs in the binary. Variability at 10.5\,$\mu$m could either be caused by ammonia ($\rm NH_{3}$) or silicates. Since the $\rm NH_{3}$ absorption features, which were first-time identified in the mid-infrared spectra taken with Spitzer/IRAS by $\cite{roellig04}$, and $\cite{cushing06}$, appear at roughly the L/T transition it is unlikely to be the cause of the observed variability. 
A more favourable explanation may be an inhomogeneous distribution of silicate clouds, which characterise the atmospheres of L dwarfs with effective temperatures of roughly 1400\,-\,2000\,K ($\cite{burrows01}$). Future VISIR observations of HD\,130948\,BC at 10.5\,$\mu$m over different timescales should secure this photometric variability.
\section{Summary}
Using VISIR at the VLT we performed a mini-survey of brown dwarfs in binary systems. The four selected brown dwarfs were imaged in three narrow band filters at 8.6, 10.5 and 11.25\,$\mu$m. At 8.6\,$\mu$m three of the brown dwarfs were detected and photometry was obtained. None of the brown dwarfs was detected at 11.25\,$\mu$m and only, HD\,130948\,BC, at 10.5\,$\mu$m. The observations of HD\,130948\,BC at 10.5\,$\mu$m indicate a possible variation of one or both brown dwarfs of the binary.

To constrain the atmospheric properties of the brown dwarfs we compared the mid-infrared photometry to theoretical model spectra by $\cite{allard01}$ and $\cite{burrows06}$. The measured mid-infrared fluxes and upper limit, respectively, of GJ\,229\,B are consistent with the characteristic parameter obtained by $\cite{saumon00}$ ($\rm T_{eff}$\,$\sim$\,950\,K, log\,{\it g}\,$\sim$\,5), while values of the effective temperature and gravity as suggested by $\cite{leggett02}$ ($\rm T_{eff}$\,$\sim$\,1000\,K, log\,{\it g}\,$\le$\,3.5) result in too high model fluxes. As for HD\,130948\,BC, the model fluxes for $\rm T_{eff}$\,$\sim$\,1900\,K, log\,{\it g}\,$\le$\,5) fit the measurement at 8.6\,$\mu$m and the upper limits obtained at 10.5\,$\mu$m and 11.25\,$\mu$m. Nevertheless the models are not in agreement with the flux measured for the detection of HD\,130948\,BC at 10.5\,$\mu$m during one observing epoch.
%%%%%%%%%%%%%%%%%%%%%%%%%%%%%%%%%%%%%%%%%%%%%%%%%%%%%%%%%%%
%%%%%%%%%%%%%%%%%%%%%%%%%%%%%%%%%%%%%%%%%%%%%%%%%%%%%%%%%%%

%
%

%
\end{document}